\documentclass[a4paper]{article}
\pdfoutput=1       

\usepackage{hyperref}

\usepackage{amsmath}
\usepackage{amsfonts}
\usepackage{amssymb}
\usepackage{graphicx}
\usepackage[section]{placeins}

\usepackage{listings}
\usepackage{caption}
\usepackage{color}
\def\cpu{{\scshape cpu}}
\def\cpus{{\scshape cpu}s}
\def\gcc{{\scshape gcc}}

\def\gpus{{\scshape gpu}s}
\def\nvidia{{\scshape nvidia}}
\def\ompi{{\scshape omp}i}

\begin{document}

\title{Experiences with task-based programming using cluster nodes as OpenMP devices}

\author{%
Ilias Keftakis\\
\textit{Dept.~of Computer Science and Engineering} \\
\textit{University of Ioannina}\\
Ioannina, Greece\\
ikleftakis@cse.uoi.gr
\and
Vassilios V.~Dimakopoulos\\
\textit{Dept.~of Computer Science and Engineering} \\
\textit{University of Ioannina}\\
Ioannina, Greece\\
dimako@cse.uoi.gr}

\date{}

\maketitle

\begin{abstract}
Programming a distributed system, such as a cluster, requires extended use
of low-level communication libraries and can often become cumbersome and 
error prone for the average developer. In this work, we consider each node 
of a cluster as a separate OpenMP device, able to run code with 
OpenMP directives in parallel. We make use of the OpenMP device model to 
provide an easy and intuitive way to program available cluster nodes. 
Based on that, we examine modifications that were necessary to make existing 
task-based applications able to exploit such an infrastructure. Finally, 
we evaluate the performance of the system and conclude that one can gain 
significant speedup, as long as the application tasks do not produce 
excessive communication overheads.
\end{abstract}

\section{Introduction}

High-performance scientific applications always require computationally 
powerful systems to execute on. On one hand, parallel systems with shared 
memory architectures are relatively easy to program, for example by using 
the directive based approach of OpenMP, but do not scale well. On the 
other hand, distributed-memory systems scale well, but programming them requires 
low-level libraries that are usually cumbersome and error prone for the 
average developer to use. As it is natural, a lot of effort has been
devoted to combine these approaches and get the best of both worlds; scalable
distributed systems that also are easy and intuitive to program. 

Early attempts tried to implement transparent execution of OpenMP programs 
on clusters using software Distributed Shared Memory (sDSM) 
libraries. The role of an sDSM library is to 
keep memory consistent between separate nodes, as it would be on a shared 
memory system. This means that whenever a global (shared) variable is modified, 
all nodes must at some point be notified about the change, in order for the 
execution of the program to proceed correctly. A prominent system to take this 
approach was Intel's Cluster OpenMP. But, even with
the help of compiler analysis to keep as many variables as possible
private to each node, these solutions did not always perform well in real 
world applications, as shown in~\cite{dsm}, because of memory sharing overheads 
and the need for global synchronization.

The fourth release of the OpenMP specification offered support for 
coprocessors and accelerators using the new \textit{target} directive. 
The programmer still 
has to manually define which parts of his code should run on the accelerator 
(or any similar ``device'') and which variables need to be sent over, but 
the implementation takes care of the low-level details concerning how 
to communicate and exchange data with the devices. 

In this work we examine the applicability of the OpenMP device model as a
means of harnessing the nodes of a cluster by an OpenMP-only application,
especially for applications whose code contains task-based computations. 
In particular:
\begin{itemize}
	\item We implement a new device module in the OMPi OpenMP compiler which
	   presents cluster nodes as separate OpenMP devices. The
     devices can execute code through the OpenMP \textit{target} directive,
     with transparent handling of code and data transfers.
	\item We elaborate on how to restructure existing task-based OpenMP 
	   applications so as to exploit such devices. That way, a computation can 
	   be distributed quite easily among the nodes of a cluster.
\end{itemize}

The rest of this paper is organized as follows. Section \ref{sec:related}
discusses related work, while Section \ref{sec:openmp} provides the 
necessary background on OpenMP and more specifically its device model.
Section \ref{sec:mpinode} thoroughly describes the design and implementation of
our new module, that is able to offload code to nodes of a cluster. Section
\ref{sec:evaluation} presents application cases
and reports on their performance. Finally, Section
\ref{sec:conclusion} concludes the paper.

\section{Related Work}
\label{sec:related}

Devices were added to OpenMP in V4.0 of its specifications 
\cite{OpenMP}. While OpenMP specifications have recently reached V5.1, 
device 
support remains rather limited; there exist relatively few compilers 
supporting relatively few device types. The Intel Xeon Phi co-processor 
can be programmed under several models and OpenMP was one of the first to be 
ported. Barker and Bowden~\cite{xeon_phi_2} measure and analyze the 
performance of two scientific applications ported to Xeon Phi, using the 
offload and native execution programming models. 

\gpus\ are arguably the devices that receive most of the OpenMP offloading 
support. The authors in~\cite{cuda} discuss their efforts implementing the 
OpenMP 4.5 specification into the Clang/LLVM project to support \nvidia\ \gpus. 
The recent versions of the \gcc\ compiler~\cite{gcc} support offloading code 
to Intel Xeon Phi accelerators and
\nvidia\ \gpus{} using the OpenMP programming model.

Envisaging remote computers as possible OpenMP devices is not a new idea,
but the related literature is quite limited.
Our approach is similar to the work by Jacob et al.~\cite{Jacob2015}. 
The authors elaborate mainly on the conceptual model 
while giving limited insights on their implementation and its functionality
within the LLVM compiler; a bioinformatics, loop-based application is then 
used to demonstrate the potential of the proposed model, experimenting 
with the loop schedules. 
In our work we give a detailed description of the device implementation 
which is available in the context of the open-source \ompi\ compiler,
and with node heterogeneity in mind.
Moreover, we utilize the new device to (re)write task-based OpenMP applications 
and report on our experiences. 

Instead of treating cluster nodes as separate 
devices, Yviquel and Araújo~\cite{8025309} propose a method to handle an entire 
cloud infrastructure as a single device using map-reduce Spark nodes and 
remote communication management. 
Finally, in \cite{clusterws2} the authors utilize the device model of OpenMP as an 
annotation mechanism to denote possibly migratable computation among 
cluster nodes. The authors do not aim to use cluster nodes as OpenMP devices; 
they rather extend their previous work \cite{clusterws1} which targets hybrid MPI+OpenMP 
applications with the aim of providing work-stealing functionality across MPI 
rank boundaries at the application level.

\section{The OpenMP Device Model}
\label{sec:openmp}

Since the introduction of OpenMP V4.0, offloading portions
of code to devices (typically \gpus\ and other accelerators) is made possible by 
using the
\textit{target} directive. Statements that are inside the block that follows the
\textit{target} directive get executed at the specified device (or at the
default device if none is specified) and not at the main processor 
(also called \textit{the host}); such blocks are
called \textit{kernels}. Other OpenMP
directives and parallel regions are allowed to exist inside kernels. The
\textit{map} clause of the directive allows the programmer 
to define a map between variables
of the host system and the device, meaning that data can be transferred 
to and from a device with one of the following four ways:
\begin{itemize}
	\item If the map type is \textit{to} the value of the variable gets copied from
the host to the device before the execution of the block. This is useful for
initializing data.
	\item If the map type is \textit{from} the value of the variable gets copied
from the device to the host after the execution of the block. This is useful
for getting the results back.
	\item If the map type is \textit{tofrom} we achieve the combination of the
previous two options. This is useful if we have data whose value will be used
and modified and we need the final value.
	\item If the map type is \textit{alloc} a data mapping between the host and the
device is created, but no data is transferred. This is useful for auxiliary
data that does not have a predefined initial value, neither we need its final
value.
\end{itemize}

To make the above points more clear, Listing \ref{prg:add} gives a simple 
example. The code performs array addition at a device. At first, variables
\texttt{a}, \texttt{b}, \texttt{c} and \texttt{size} exist only in host's
memory. When the \textit{target} directive gets executed, memory is allocated at
the device for the arrays \texttt{a}, \texttt{b} and \texttt{c} of \texttt{size}
elements each, starting at position \texttt{0}, because of the
\texttt{[0:size]} section. Memory for the variable \texttt{size} is also 
allocated. The
values of the arrays \texttt{a} and \texttt{b}, as well as the variable
\texttt{size} which are mapped with type \textit{to} get copied from host's to
device's memory.

\begin{lstlisting}[float=tb,label=prg:add,caption=Array addition at a device.]
double a[1024];
double b[1024];
double c[1024];
int size = 1024;

void add_at_device(double *a, double *b, 
                   double *c, int size)
{
	#pragma omp target map(to:a[0:size],b[0:size],size) map(from: c[0:size])
	{
		int i;
		#pragma omp parallel for
		for (i = 0; i < size; i++)
			c[i] = a[i] + b[i];
	}
}
\end{lstlisting}

Subsequently, the array addition gets offloaded and executed in parallel 
at the device,
because of the \textit{parallel for} OpenMP directive. That way one can fully
utilize the device's processing cores. When the calculation is completed, the
elements of array \texttt{c} which is mapped with type \textit{from} get copied
from the device's to the host's memory. It is useful to remember that 
data transfers are required given that the host and the device have different
memory spaces; in cases where the device shares memory with the host, 
some or all data may be accessed through shared memory, possibly 
avoiding copies and transfers.

\section{The mpinode Device}
\label{sec:mpinode}

The \ompi\ compiler~\cite{ompi} is a lightweight OpenMP C 
infrastructure, 
composed of a source-to-source translator and a flexible, modular runtime 
system. 
\ompi\ is an open source project and targets general-purpose SMPs and multicore 
platforms. It provides a large portion of the OpenMP V4.5 functionalities,
including full support for device constructs.

The compiler uses \textit{outlining} to move the code of each target block into 
a separate function, called a \textit{kernel function}. The final executable 
file consists of all user's code as well as all kernel functions. 

Regarding devices, the runtime system of \ompi\ is organized as a collection 
of \textit{modules} each one implementing support for a particular device class; 
multiple devices (of the same type) may be served by one module. 
Modules consist of two parts:
the host part and the device part. The former enables the host (main \cpu)
to access any of the available module's devices through a fixed interface
and is loaded on demand as a plugin (shared library). The device part
provides OpenMP and other runtime support within the device, for the
offloaded code.

To support a new device in \ompi, one has to create a new module that implements
\ompi's interface for communication with devices. For our purposes here, we
created the \textit{mpinode} module; an \textit{mpinode} device is simply a
computer with MPI installed. In order to exploit cluster nodes,
the user is required to create a simple configuration file. All nodes appear in
this file and every line contains the name or IP address of a single node,
followed optionally by an integer. Each of the listed nodes is considered 
as a separate device where the user can offload code to. Appending an 
integer, say $D$, is equivalent to having the node appear in $D$ different lines 
in the configuration file. In such a case, $D$ devices will be 
started at that node. The user may find this characteristic useful to better 
utilize a node's resources, if the offloaded codes do not generate 
enough parallelism.

\subsection{Operation}

When a program starts, the initial process running on the host 
calls \texttt{MPI\_spawn()}
to spawn a process on each node specified in the configuration file. Each
process then calls \texttt{MPI\_Intercomm\_merge()} to merge the 
communicators, so that the process running on the host gets assigned
MPI rank 0. When MPI creates a new process, it just copies the executable 
to the new node, so processes check their MPI rank to determine if they run 
at the host or at a device. The process that runs at the host is the only one 
that executes the program, including all user's code; all the other processes
first perform some necessary initialization and then enter a loop waiting 
for commands from the host.

Communication is achieved by using the MPI library. Two devices cannot
communicate with each other directly (that is without mediation of the host);
this is an OpenMP restriction. Devices wait for a command from the host, 
execute it and then return to their waiting state. 
Here are the types of commands remote \textit{mpinode} processes listen for:
\begin{itemize}
 \item Allocate or deallocate memory at the device.
 \item Transfer data to or from the device.
 \item Execute a kernel function.
 \item Stop a device when the application exits.
\end{itemize}
A command might be composed of a sequence of MPI messages; for
example, the host sends a message asking for the value of a variable and the
device replies with a message that contains it. 
The device must always know the type of the 
command it is asked to execute, so the host always includes it 
in the first message.

Because the processes operating at remote nodes have been spawned as 
replicas of the initial host process, they contain all kernel functions.
Consequently, there is no need to actually transfer the kernel code at 
offloading time. However, the host cannot instruct an \textit{mpinode} device 
to execute a kernel function by just sending the memory address of the 
function, since different machines may store program code and data at different 
virtual addresses, even if they run the same executable file. 
To solve this problem, every node constructs a
\textit{kerneltable} structure, which maps each kernel function to a
unique integer. Internally, the kerneltable is a
dynamically allocated array whose every element consists of the name of a kernel
function and a pointer to its code. At offload time, the host searches its 
kerneltable for the given kernel function name and simply sends the array index 
that corresponds to the function to the device. The device uses that index 
on its own kerneltable to retrieve a pointer (in its own address space) to 
the function it should execute.
The kerneltable is populated at program start, before executing any user code,
by sequentially inserting each kernel function's name and pointer.
This occurs independently at each cluster node and because every node runs 
the same executable, functions are entered in each kerneltable in the 
exact same order; as a result, each function is mapped to the same unique integer
across all nodes.

\subsection{Data Handling}

As per OpenMP specifications, for each \textit{target} region the host must 
maintain mappings of all host variables with the corresponding device
variables, so that transfers can be initiated to/from the host address space 
form/to the device space. In the case of cluster nodes, it may 
be difficult or even impossible for the host to know the exact virtual 
address where the corresponding (remote) variables will be stored at.
For such scenarios \ompi\ introduces a level of indirection, which
we call a \textit{mediary address}: Instead of mapping a host address to
a device address, it maps a host address to an abstract mediary address. 
Transfers to/from the device utilize the mediary addresses; the device is 
then responsible to translate mediary addresses to its own memory space 
addresses. It is up to the module implementor to decide what exactly the 
mediary address represents.
With this mechanism, the host is able to create mappings even if 
the corresponding variables have not yet been allocated on the remote device.

When handling a \textit{target} directive, the host creates a struct 
in which it places the mediary address for each variable that is accessed 
in the \textit{target} block. Then, before commanding kernel execution,
it sends the struct to the appropriate device which uses it to extract 
the actual device address of each variable.
These device addresses will subsequently be used during kernel execution.
For the \textit{mpinode} module, mediary addresses are simple integers and are 
implemented as follows:
\begin{itemize}
\item Each device keeps a dynamic array which maintains all memory allocation
      requests. When the host asks for a new allocation (e.g.\ 
      requesting space for a corresponding variable), the first unused 
      array element is found and the requested amount of memory is allocated
      on the device with a \texttt{calloc()} call, storing its address at 
      this array element. Unused elements contain the \texttt{NULL} address.
      The address stored at the array element is the 
      actual device address of the corresponding variable, while the array 
      index of the element is used as the mediary address of the variable;
      the latter is used when communicating with the host.
\item The host is not technically required to keep such an array; it could ask
      the device for the next available mediary address, but that would 
      require further message exchanges and thus cause additional overheads. 
      Consequently, as an optimization, the host maintains a mirror of the 
      above array for each device. Of course, the host does not need to 
      allocate any memory, it only needs to remember which elements are in
      use. Accordingly, to allocate memory for a new variable on a device, 
      the host first finds the first unused array element and marks it with 
      the special (and arbitrary) value of \texttt{0x999}. Then it notifies 
      the device to proceed with allocating memory, as described above. 
      This way the host knows in advance the mediary address that the device
      will use. Unused elements again contain the \texttt{NULL} address.
\end{itemize}

All nodes place the addresses of global variables (such as variables 
declared within \textit{declare target} regions) in their arrays at the 
beginning of the execution and in the same order, before executing any user 
code. Global variables and their addresses are retrieved from \ompi's internal 
data structures.

\begin{figure}[tb]
	\centering
	\includegraphics[width=0.7\columnwidth]{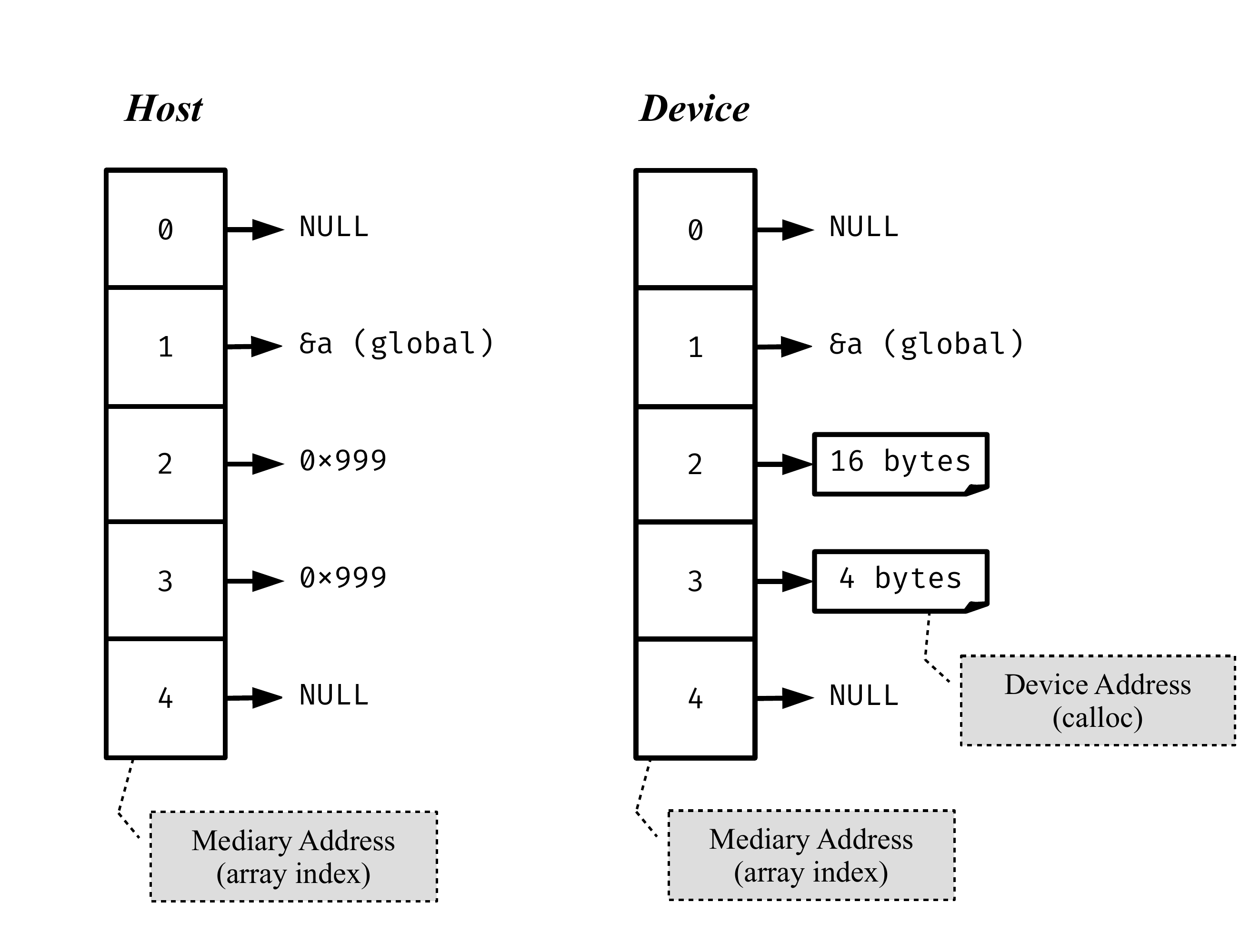}
	\caption{Example of a mediary address array at the host (left) and 
	the device (right).}
	\label{fig:address}
\end{figure}

Figure \ref{fig:address} shows an example of a mediary address array at the 
host (on the left) and a device (on the right). We suppose that the example 
program only has one global variable named \texttt{a}. 
Also, the host has asked the device
to allocate memory two times; the first one for an array of 16 bytes and the
second for a variable of 4 bytes. Naturally, when a \textit{target} block is
done executing and any required data transfers have taken place, allocated
variables are freed from the device's mediary address array and their positions
are marked as unused at both the host and the device array.

For the best possible performance, one may employ OpenMP \textit{target}
directives inside \textit{parallel} directives, meaning that different
threads at the host can communicate with different devices simultaneously. 
That requires the \textit{mpinode} module to be thread safe; this in turn requires
the MPI implementation to be thread safe. Furthermore, many
communication operations require two phases; first the host sends data to the
device and then calls \texttt{MPI\_Recv()} to get the response. If two threads
at host communicate with the same device simultaneously, one might end up
receiving the other's data. If they communicate with different devices, the
problem does not occur because the host checks the source rank of the device
process when calling \texttt{MPI\_Recv()}. We could use the \textit{tag}
parameter of \texttt{MPI\_Recv()} calls to solve the problem, but we decided to
have it available for future use. Instead, every time we want to use MPI
functions, we check if we are inside a parallel region and, if so, we lock a
\textit{mutex} dedicated to the device we want to use. When the host inserts or
removes entries from the mediary address array, it also uses the same mutex to
prevent data races at the array.

\subsection{The Device Part}

As mentioned previously, the device part of an \ompi\ module is responsible 
for providing OpenMP facilities and runtime support to an executing kernel.
In common cases it has the form of a static library that gets linked with 
the offloaded kernel, implementing the required services. 
In our case, the device part of the \textit{mpinode} module is effectively 
empty. The reason is that the executable that runs across the cluster 
already contains the host's runtime library linked in. That is, a fully
fledged OpenMP runtime is already present in the executable and is utilized
as-is to offer complete OpenMP facilities to any kernel executing on any 
remote node.

\section{Evaluation}
\label{sec:evaluation}

\subsection{Methodology}
Once the \textit{mpinode} module is implemented, every node in a cluster
is accessible as a separate OpenMP device.
The proposed system represents a straightforward way to exploit multiple
cluster nodes without resorting to complex message passing. It can be used 
to easily (sometimes trivially) offload any kind of code to cluster nodes 
using only OpenMP directives. Conceptually, a block of code accompanied by 
the required data are sent for execution to other \cpu{}s; this aligns 
perfectly with the tasking model of OpenMP and opens up the possibility of
programming a whole cluster by only using OpenMP tasks.

To assess the programmability as well as evaluate the performance of such 
an approach, we use selected applications from the Barcelona OpenMP Task Suite
(BOTS)~\cite{5361951}, and more specifically alignment, fib and sparselu. All
of them use OpenMP tasks to perform parallel calculations on a
shared-memory machine. We modified them by adding \textit{target} directives 
to execute portions of the calculations at different nodes. Apart from BOTS, 
we also parallelized a sequential Mandelbrot set application using OpenMP 
tasks and our \textit{mpinode} module.

The general idea when modifying existing applications is to find computationally
intensive parts of the code that do not require a lot of data for their
calculations and wrap them inside \textit{target} directives. Since a node of a
typical cluster consists of more than one \cpus\ and many processing cores, it is
beneficial if the \textit{target} block contains additional \textit{parallel
for} or \textit{task} directives, so the node's resources are better utilized.
Alternatively, the same node can be treated as multiple different devices by
including its IP address multiple times in the configuration file. 

In some
cases, for example when having nested \textit{task} directives, all one has to
do is rename the external \textit{task} into \textit{target}. But, most of
the time, it requires more work; one needs to create as many \textit{target}
regions as there are devices (for example by splitting the iterations of a loop)
and specify which variables and which array sections each device will use. In
any case, for the average user the process is faster, easier and less error 
prone than having to restructure a program to utilize MPI.

\begin{lstlisting}[float=tb,label=prg:addmulti,caption=Array addition at multiple devices.]
double a[1024];
double b[1024];
double c[1024];
int ndevices = 8;

#define START device*size
#define END   device*size+size

void add_multidev(double *a, double *b,
                  double *c, int ndevices)
{
	int size = 1024 / ndevices;
	int device;

	#pragma omp master
	for (device = 0; device < ndevices; device++)
	{
		#pragma omp target device(device+1) map(from:c[START:END]) map(to:a[START:END],b[START:END]) firstprivate(size,device) nowait
		{
			#pragma omp parallel for
			for (int i = START; i < END; i++)
				c[i] = a[i] + b[i];
		}
	}
}
\end{lstlisting}

As a concrete example, in Listing \ref{prg:addmulti} we revisit 
Listing \ref{prg:add}, which performed array addition
at a single device; the new program utilizes 8 devices to perform the
addition. We assume that the \texttt{add\_multidev} function 
is called from a parallel region. The loop in line 16 creates a \textit{target} region for each
of the 8 devices. Since we want this operation to be performed once, we use a
\textit{master} directive in line 15.

The \textit{target} directive at line 18 utilizes the device specified by the
\textit{device} clause. Note that we start from device 1, since device 0 is the
host machine. As explained in Section \ref{sec:openmp}, arrays \texttt{a} and
\texttt{b}, as well as variables \texttt{size} and \texttt{device} are copied to
the device, while array \texttt{c} is copied back to the host when execution
reaches line 23. In contrast to Listing \ref{prg:add} where the entire arrays
are copied, only the required 128 elements of each array are copied per device,
using appropriate array sections. 
The \textit{nowait} clause indicates that the host thread will not wait until
the offloaded region is completed. To fully utilize the cores within each 
device, we create a \textit{parallel for} region in line 20.

\subsection{System Specifications}

All experimentation took place at a commodity cluster we had direct access to. 
The cluster consists of 16 nodes with 2 \cpus\ each (dual core
AMD Opteron @ 2193MHz) and 12GB memory. Nodes communicate through a 
humble Gbit Ethernet network. Each node runs Ubuntu 16.04.4 LTS with 
Linux 4.4.0 and uses \gcc\ 5.4.0 and OpenMPI 3.0.2. 
To evaluate the performance 
of our module, we used 2 to 12 nodes, depending on the benchmark. We measured 
the speedup we gained compared to execution in one machine 
(where we executed the original version of each benchmark using only a 
single random node of the cluster).

For an unbiased comparison, we used \gcc\ as the reference OpenMP 
implementation instead of the original \ompi. However, the problem we faced
is that the tasking implementation in the available version of \gcc\ performs 
inexplicably poorly in some cases, resulting in high execution times and
false superlinear speedups for our module. Because of such cases, we also 
report speedups with respect to using our module on a single machine; this 
also serves to show how \textit{mpinode}
performance scales with the utilization of more nodes.
We executed each experiment 10 times and averaged the results.
We note that we did not alter the way BOTS measure and report performance 
results, i.e.\ they measure the execution time (wall clock time) of the 
parallel section of the program, not the time required for the initialization 
or the finalization. Thus, the reported times do not include the (one-shot) 
delays of starting and stopping MPI processes on remote nodes.

\subsection{Protein Alignment}

The alignment program performs protein alignment. 
From a programmer's perspective,
it produces a one dimensional array, where each element can be calculated
independently of all the others. For its calculation, many arithmetic
operations with auxiliary arrays and variables are required, which do not change
during the execution of the program, and thus can be sent once at each device
(at the beginning of the execution). We modified this program so that 
if we have $n$
devices and the output array has $m$ elements, each device calculates $m/n$
elements. We were careful to send to devices only the portions of the arrays
that were absolutely necessary to calculate their assigned work, in order to
limit the communications as much as possible.

As a result, we achieve parallelization at two levels. First, the host separates
the work into equal parts and simultaneously asks the devices to calculate their
part (using a \textit{target} directive). Secondly, each device executes its
kernel function in parallel by creating multiple threads to calculate the result
it was assigned (using a \textit{task} directive).

\begin{figure}[tb]
	\centering
	\includegraphics[width=.75\columnwidth]{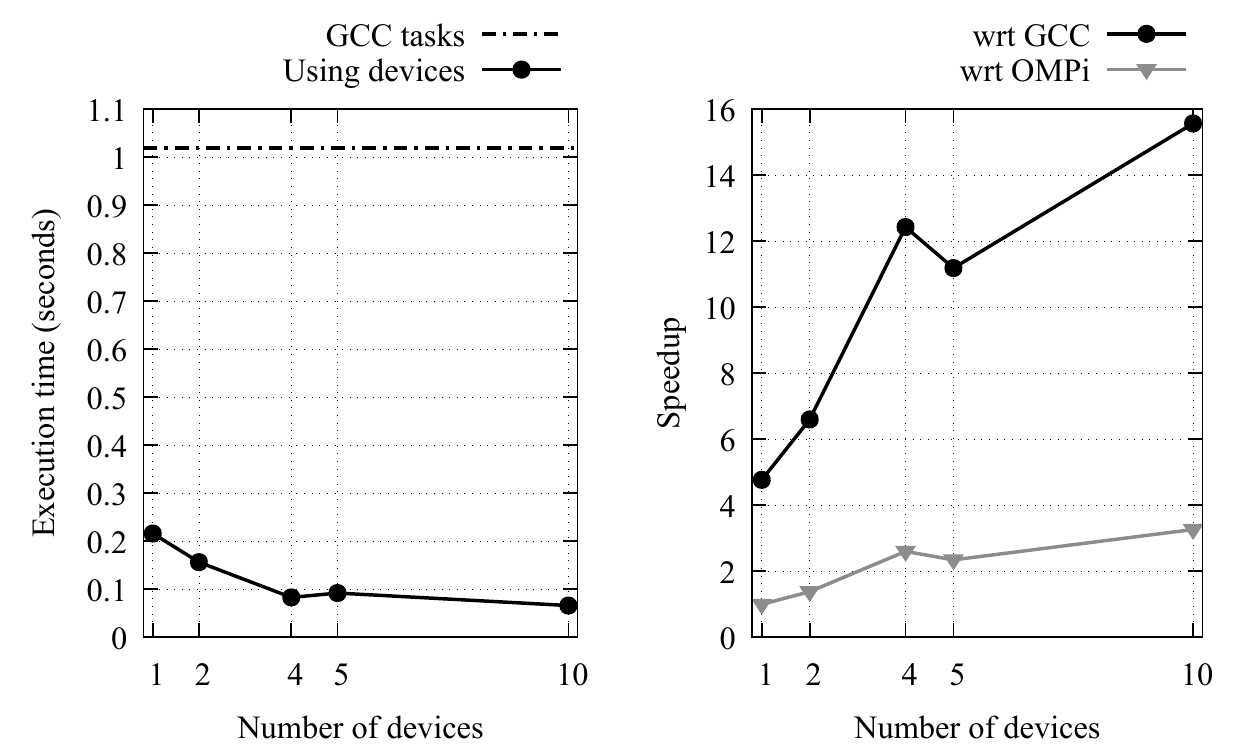}
	\caption{Plot of execution time and speedup of alignment program for input size
	20.}
	\label{fig:align20}
\end{figure}
\begin{figure}[tb]
	\centering
	\includegraphics[width=.75\columnwidth]{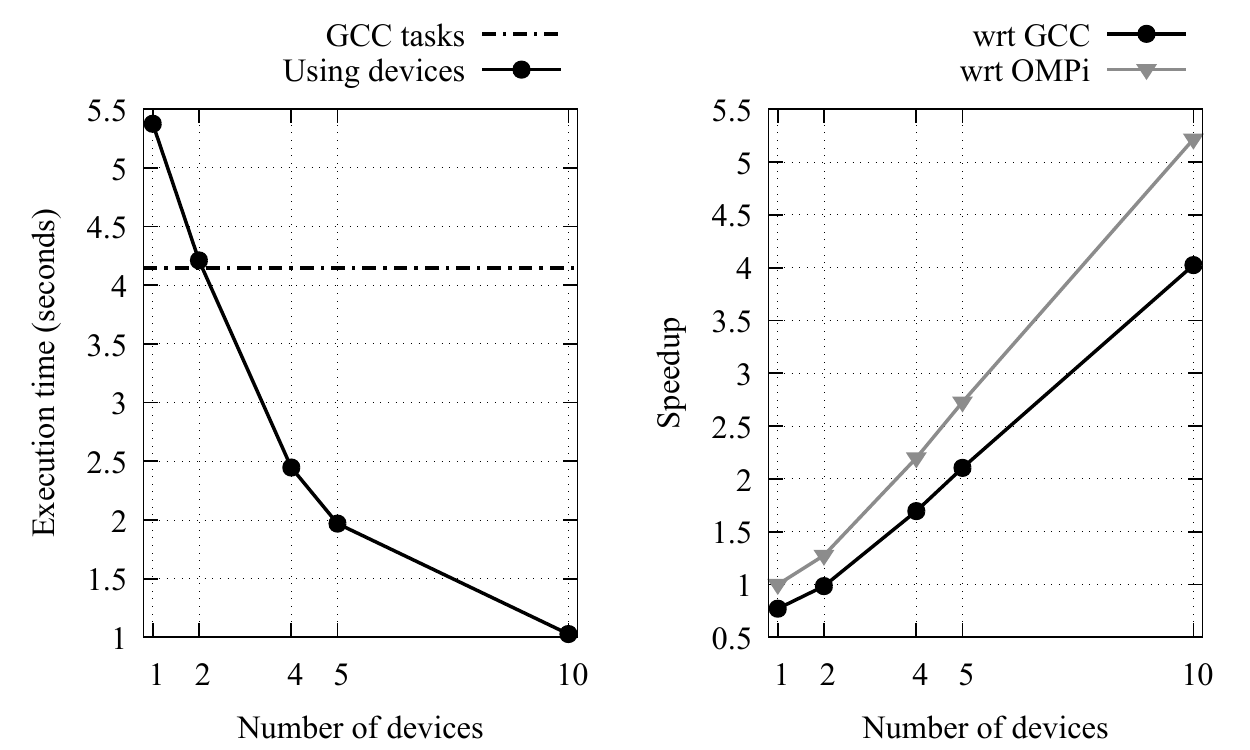}
	\caption{Plot of execution time and speedup of alignment program for input size
	100.}
	\label{fig:align100}
\end{figure}

Figures \ref{fig:align20} and \ref{fig:align100} show the execution time of the
alignment program, when given as input the 
files \texttt{prot20.aa} (that produces
an output array of 400 elements) and \texttt{prot100.aa} (that produces an
output array of 10000 elements), respectively. The execution time when using
only OpenMP \textit{tasks}, that uses 4 threads in one node, is shown with
dashed lines. In all other cases, the total number of threads we use is 4
multiplied by the number of devices. The communication overhead is small, since
the amount of communicated data is small, but each device has quite 
a lot of work to
do. Furthermore, the biggest part of communication takes place once, at the
beginning of the program. For the \texttt{prot20.aa} file, 
\gcc\ was problematic, exhibiting consistently high execution times.
As a result, the speedup we achieve reaches the unreal value of 15.57. 
For the larger problem of \texttt{prot100.aa}
our system scales linearly with the number of nodes.

\subsection{Mandelbrot Set}
The mandelbrot program creates a Mandelbrot fractal image (which is a two
dimensional array) with the provided dimensions.
The calculation of the value of each pixel requires only
the position $(x, y)$ of the pixel. We parallelized a sequential version of
the program such that each
device constructs a strip (a number of adjacent rows) of the image, so the only
information a device needs is the position and the size of the strip it has to
construct. We also added a \textit{parallel for} directive so that each device
can calculate its rows in parallel. When a strip is completed, it needs to be
sent back to the host, and that can require a significant amount of
communication, depending on the image dimensions.

\begin{figure}[tb]
	\centering
	\includegraphics[width=.75\columnwidth]{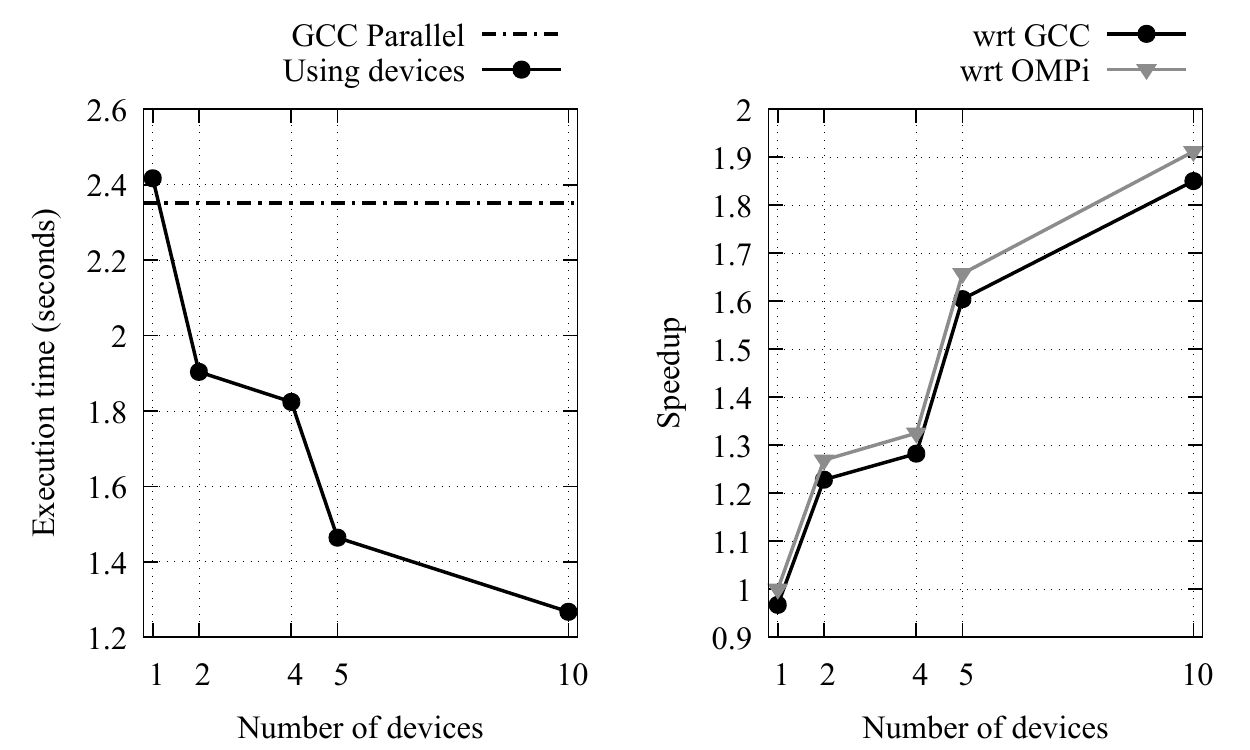}
	\caption{Plot of execution time and speedup of mandelbrot program for an image
	with size 2600x2600 pixels.}
	\label{fig:mandel2600}
\end{figure}
\begin{figure}[tb]
	\centering
	\includegraphics[width=.75\columnwidth]{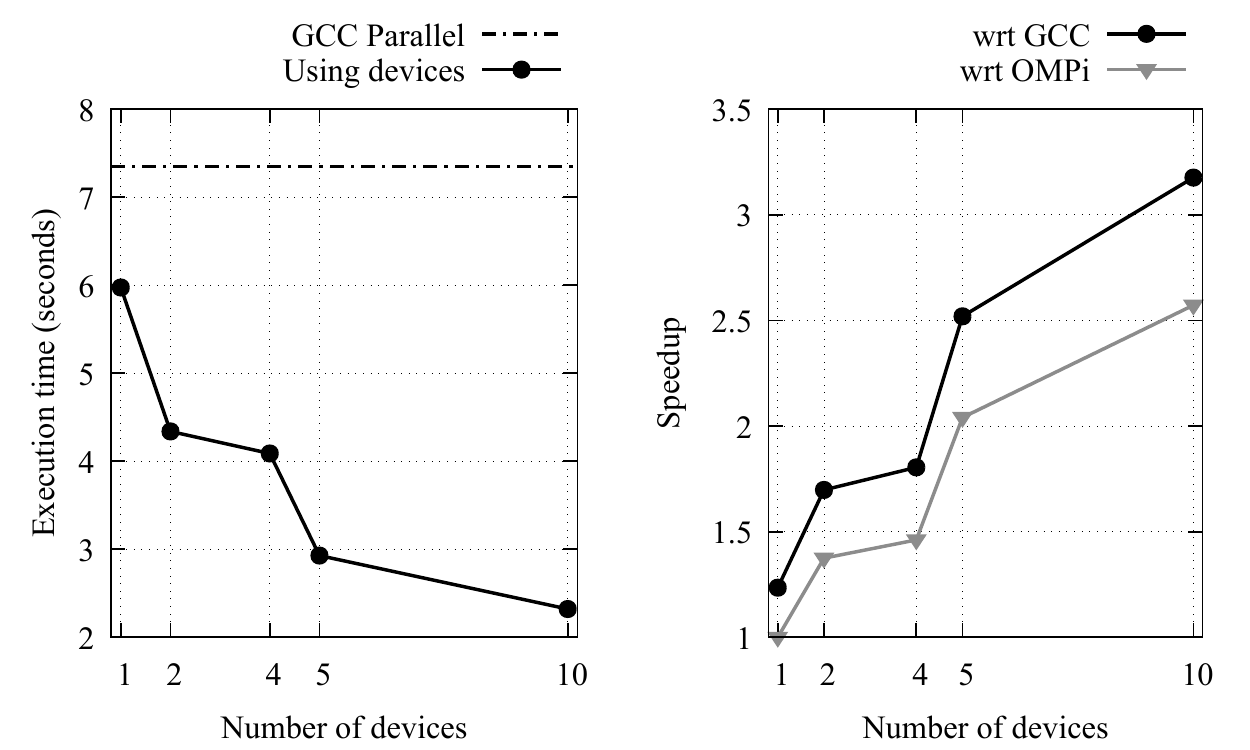}
	\caption{Plot of execution time and speedup of mandelbrot program for an image
	with size 4600x4600 pixels.}
	\label{fig:mandel4600}
\end{figure}

Figures \ref{fig:mandel2600} and \ref{fig:mandel4600} show the execution time of
the mandelbrot program, when generating an image of 2600x2600 and 4600x4600
pixels, respectively. In the former case, the speedup we gain is 1.85, while in
the latter it is 3.18, with respect to \gcc. 
We observe that when doubling the image dimensions, the
work load increases significantly (the sequential execution time triples), 
but the amount of communications does not increase as dramatically, 
thus resulting in increased speedup.

\subsection{Fibonacci}
The fib program calculates a given Fibonacci number by performing only recursive
calls. Because of the recursion we needed to adapt our strategy. The limitation
is that OpenMP only allows communication between the host and a device, not
between two devices, meaning that a device cannot send work to another one.
Thus, if the root task is offloaded to a cluster node, all subsequent 
recursive tasks will have to be executed in the same node.
To solve this problem, we let the host execute the first recursive calls.
When the
recursion unwinds to the point where the number of recursive calls 
(i.e.\ generated tasks) is equal to the number of available devices, the host 
can offload the tasks to the devices and wait for their results. Notice, however,
that there is an inherent imbalance here as the generated tasks do not contain 
the same amount of computations. Each device 
employs multiple threads to execute the tasks that were present 
in the original version of the program in parallel. The only communication 
required for the program is that the host must
send a single integer to each device (that represents the Fibonacci number the
device has to compute) and receive a single integer from each device (that
represents the computed result).

\begin{figure}[tb]
	\centering
	\includegraphics[width=.75\columnwidth]{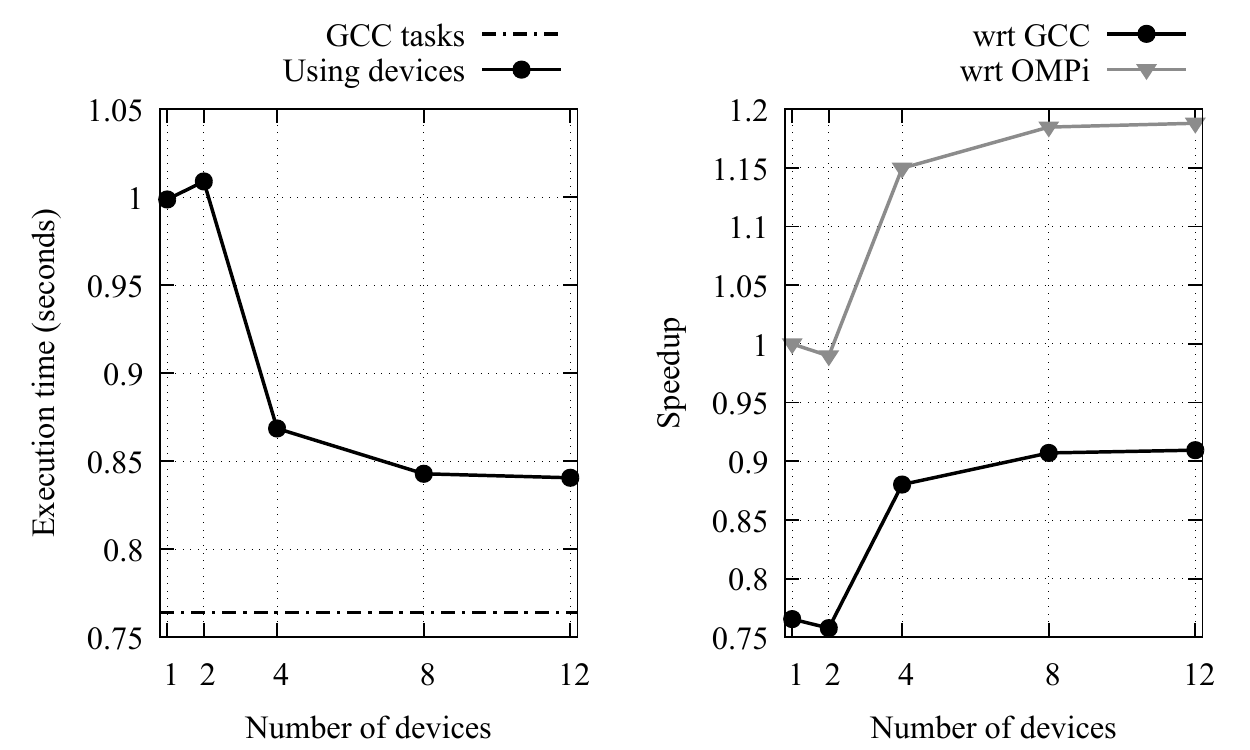}
	\caption{Plot of execution time and speedup of fib program for the number 35.}
	\label{fig:fib35}
\end{figure}
\begin{figure}[tb]
	\centering
	\includegraphics[width=.75\columnwidth]{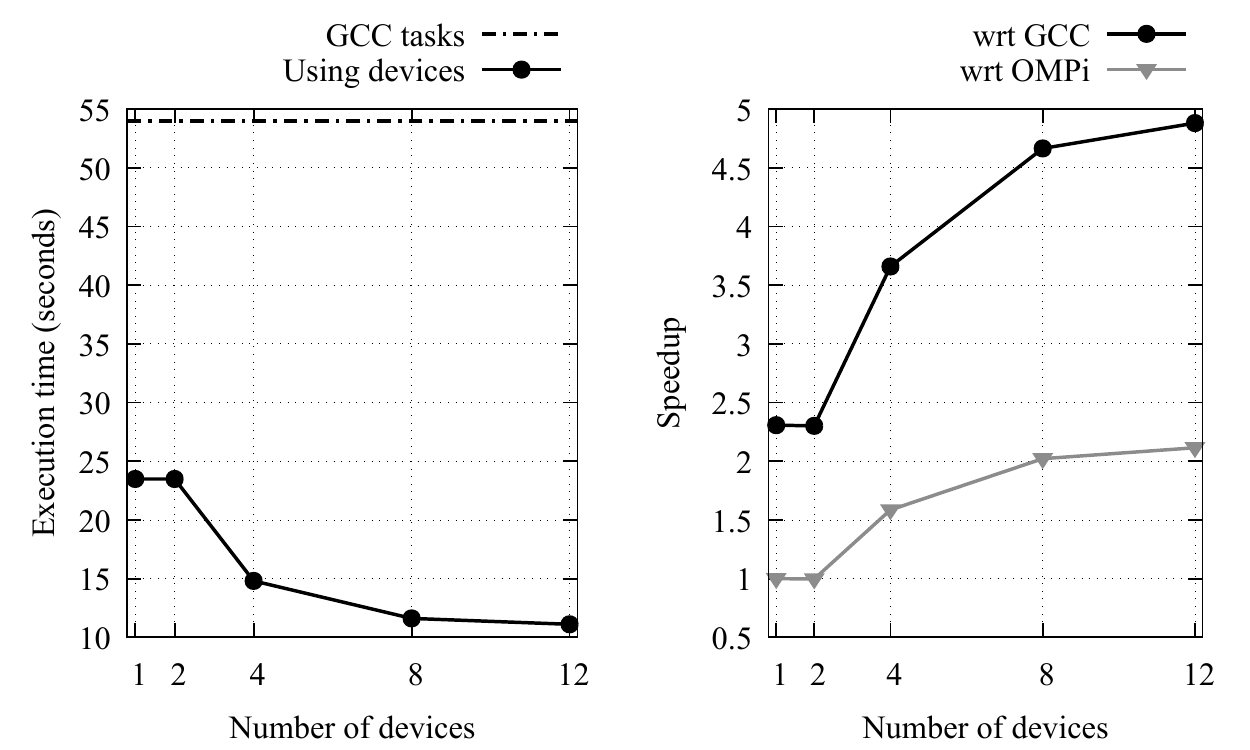}
	\caption{Plot of execution time and speedup of fib program for the number 45.}
	\label{fig:fib45}
\end{figure}

Figure \ref{fig:fib35} shows the execution time of the fib program, for the
calculation of Fibonacci number 35. We observe that the calculation of this
number at our cluster requires less than one second. The speedup we gain
compared to using only \textit{tasks} at one node reaches 0.91 in the best
case; this implies that there is not enough work to keep nodes busy, and
using \textit{task}s in a single node gives better performance.
On the other hand, the calculation of 45th Fibonacci number produces deeper 
recursions and a huge number of tasks, requiring significantly more 
computation time and producing less imbalance. 
As a result, as shown in Figure \ref{fig:fib45}, 
the addition of more devices seems to produce small but 
not negligible speedups; the large value of 4.88 with respect to \gcc\ is
due to its suboptimal tasking performance.

\subsection{Sparse LU}
Finally, the sparselu program performs LU analysis on a sparse array, whose
dimensions are given by the user. We modified that program so that each device
processes only a portion of the array in parallel, using the \textit{task}
blocks of the original version of the program. 
As a result, the host separates the array
into smaller ones of equal size and sends a sub-array to each device.
Furthermore, when the calculation is completed, the resulting sub-arrays must be
sent back to host. That implies large communication volumes that get worse 
as the array dimensions increase; in essence, the whole array must be 
transferred two times.

\begin{figure}[tb]
	\centering
	\includegraphics[width=.75\columnwidth]{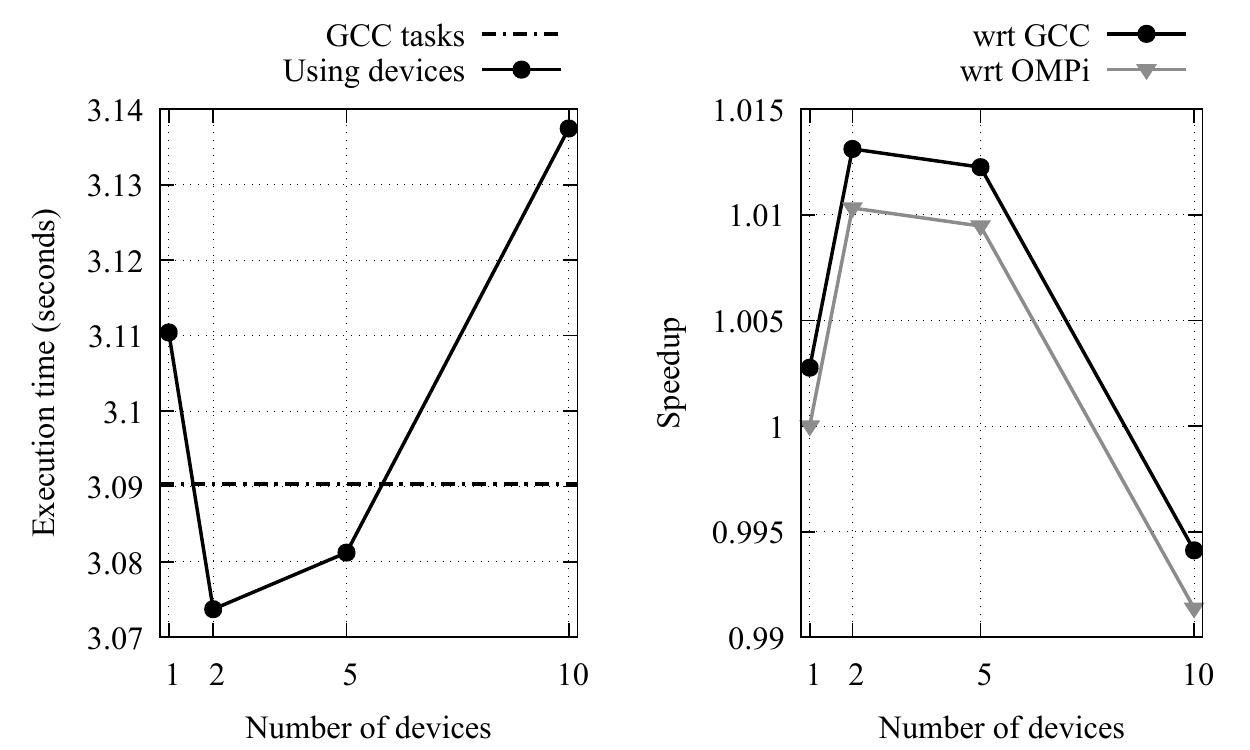}
	\caption{Plot of execution time and speedup of sparselu program for an array of
	size 2500x10000.}
	\label{fig:sparselu2500}
\end{figure}
\begin{figure}[tb]
	\centering
	\includegraphics[width=.75\columnwidth]{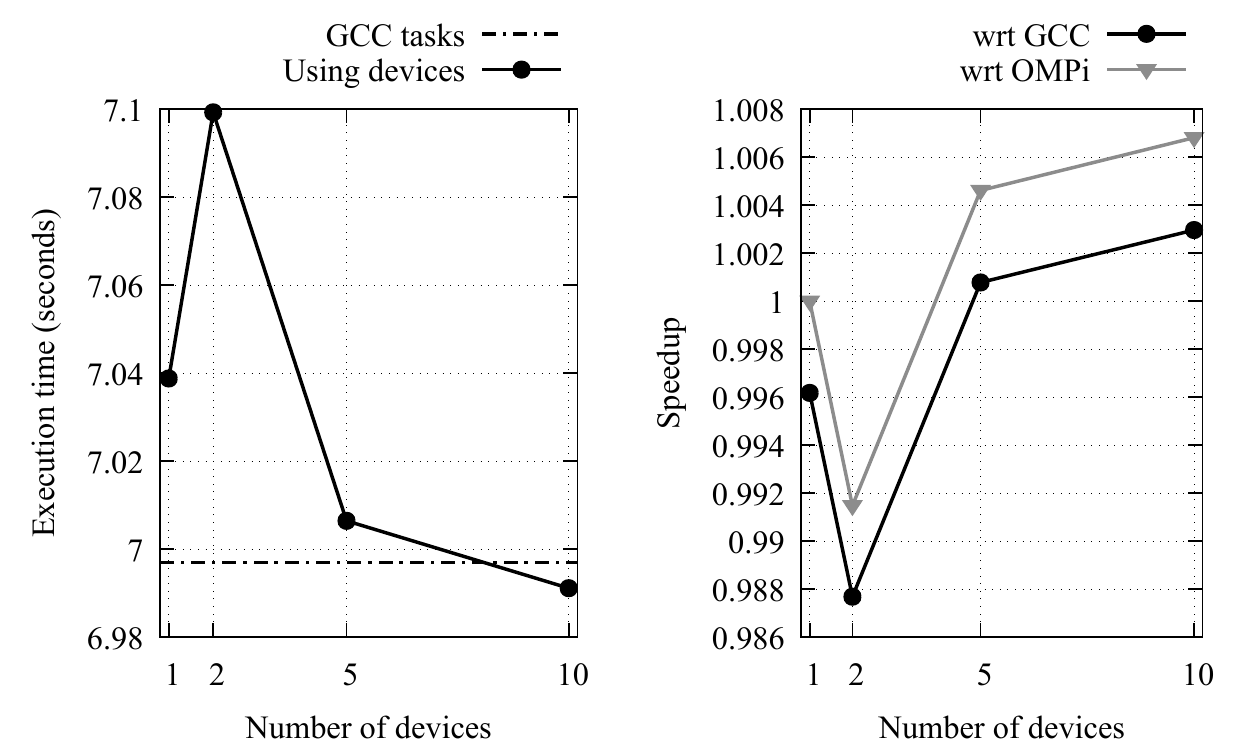}
	\caption{Plot of execution time and speedup of sparselu program for an array of
	size 3600x14400.}
	\label{fig:sparselu3600}
\end{figure}

Figures \ref{fig:sparselu2500} and \ref{fig:sparselu3600} show the execution
time of the sparselu program, for arrays of size 2500x10000 and 3600x14400,
respectively. Because of the serious communication delays, we 
essentially gain no speedups. Using cluster offloading
for applications that require a lot of communication, does not seem
a very beneficial idea in general.

\section{Conclusion}
\label{sec:conclusion}

In this paper we present the design and implementation of a module
that treats each node in a cluster as a separate OpenMP device, 
capable of running parallel code that contains arbitrary OpenMP constructs. 
Without resorting
to message passing, this facility represents a very simple way for an
OpenMP application to extract additional computational power from a cluster, 
when available, essentially for free.

We present the implementation details of our module in the context of
the \ompi\ compiler, including the way we use MPI to
handle the communication between the host and the devices and the way we 
assign mediary addresses to variables that need to be transferred to devices.

Finally we provide our experiences with utilizing the new module as a means to
deploy task-based OpenMP applications onto clusters. We analyze the 
programming techniques employed when modifying existing OpenMP applications 
to use \textit{target} constructs and kernel functions. Our experimentation
shows that with a very small effort one can gain 
significant speedups, as long as the application does not require
extensive communication between the host and the devices and each device has
enough computations to perform.

The need for intensive communication between the host and the
devices is the main source of performance degradation. We currently 
analyze and profile the communication patterns, in order to find ways to 
reduce the overheads, or avoid them altogether, if possible. Furthermore, 
we plan to extend OpenMP directives to provide a more user friendly way of 
partitioning arrays and sharing them among devices. This
way, it may also be possible to use MPI collective communications to
optimize communication between host and devices.

The \textit{mpinode} module is available in the
official repository of \ompi\ (\url{https://paragroup.cse.uoi.gr/wpsite/software/ompi/}).

\small
\frenchspacing
\bibliographystyle{ieeetr}
\bibliography{paper}
\nonfrenchspacing

\end{document}